\documentclass[11pt]{article} 

\usepackage{color}
\usepackage{authblk}
\usepackage{amsmath}
\usepackage[title]{appendix} 
\usepackage{url}
\usepackage{subfig}
\usepackage{booktabs}
\usepackage{caption}

\title{Generating long-horizon stock ``buy" signals with a neural language model }
\author{Joel R. Bock \\
New Braunfels, TX, USA}

\begin{document}

\maketitle

\begin{abstract}
	This paper describes experiments on fine-tuning a small language model to generate forecasts of long-horizon stock price movements. Inputs to the model are narrative text from 10-K reports of large market capitalization companies in the S\&P 500 index; the output is a forward-looking \textit{buy} or \textit{sell} decision. Price direction is predicted at discrete horizons up to 12 months after the report filing date. The results reported here demonstrate good out-of-sample statistical
	performance (F1-macro= 0.62) at medium to long investment horizons. In particular, the \textit{buy} signals generated from 10-K text are found most precise at 6 and 9 months in the future. As measured by the F1 score, the \textit{buy} signal provides between 4.8 and 9 percent improvement against a random stock selection model. In contrast, \textit{sell} signals generated by the models do not perform well. This may be attributed to the highly imbalanced out-of-sample data, or perhaps due to management drafting annual reports with a bias toward positive language.
	Cross-sectional analysis of performance by economic sector suggests that idiosyncratic reporting styles within industries are correlated with varying degrees and time scales of price movement predictability.
\end{abstract}

\section{Introduction}

Stock price forecasting is of fundamental interest to traders, investors, analysts and researchers motivated by the desire to increase the rate of wealth accumulation. The computational finance literature is continuously updated with studies applying the latest technological tools and methods to search for excess returns in pursuit of this objective. Large language models (LLMs) based on the transformer architecture \cite{Radford2018} have enabled the integration of nearly unlimited textual data from multiple sources along with other features derived from fundamental or technical analyses to improve financial decision making. 

Publicly-traded firms produce annual reports describing their business activities and financial statements covering the previous fiscal year. The information in these reports is also filed with the Securities and Exchange Commission (SEC) as Form 10-K. Typically this filing contains more detailed information than is found in the annual report, to include identification of market risks, legal proceedings, and other data \cite{Securities2024}. Scrutinizing all sections of 10-K report provides an in-depth picture of the company's ongoing viability and earnings growth potential. This is essential reading for analysts and investors in the company.

Language models have shown a surprising ability to learn statistical relationships between words, sentences and concepts contained in large, unstructured documents. In the present study, a pre-trained language model is fine-tuned \cite{Radford2018} to generate forecasts of future stock price movements. Inputs to the model are raw text from 10-K reports of large market capitalization companies in the S\&P 500 index; the output is a forward-looking \textit{buy} or \textit{sell} decision. No structured financial input data are used; numeric data enters only as it appears in the narrative sections of the report. Stock price movements are predicted at discrete horizons up to 12 months after report filing.

The main contributions of this paper are as follows:

\begin{itemize}
	\item The results reported here demonstrate good out-of-sample statistical performance (F1-macro$=0.62$), at medium to long investment horizons\footnote{In the present context, a ``long horizon" is taken to mean several months  to  one year.}. In particular, the ``buy" signals generated from 10-K text are found most precise at 6 and 9 months in the future;
	
	\item A cross-sectional analysis of performance by economic sector suggests that characteristic reporting styles within industries are correlated with varying degrees and time scales of price movement predictability;
	
	\item Favorable statistical performance in comparison to the existing literature, using only text in the absence of additional structured financial data to train the language models;
	
	\item It is demonstrated that relatively small language models (60M parameters) running on  a desktop machine are sufficient to achieve useful forecasting results.
	
	\color{black}
\end{itemize}

\subsection{Related work}
Copious research works are found in the literature that apply LLMs to stock price or earnings prediction. The most frequent application studied is short-term trading, with horizons on the order of days or weeks. Strong predictive accuracy has been reported by using multi-modal data from tweets, news headlines and financial time series to predict stock prices for the next trading day. Examples of this include \cite{Koa2024}, \cite{Deng2024}, \cite{Tong2024} and \cite{Wang2024}, with reported accuracies ranging from 54\% to 66\%. At slightly longer holding periods of 30 days, continuously-rebalanced portfolios of S\&P 100 stocks were documented in \cite{Fatouros2024}. The authors report ``buy" signal win-rates on the order of 66\%.

Prediction of prices at longer time frames would ostensibly seem to be a more difficult objective. In the medium to long term, models may not benefit from simple price momentum, earnings reports or other news headlines. Furthermore, unforeseen exogenous events can occur over the investment period and impact stock prices at broad horizons.

Several studies covering longer term forecasting inform and provide a basis for comparison to the present work. In \cite{Kim2023}, multi-modal interactions between 10-K textual and financial data were investigated using a large language model. To predict future earnings, 10-K text were encoded as features and input to an artificial neural network. Good accuracy and F1-macro scores (0.61) on directional movement of earnings for the ensuing fiscal year were reported \cite{Kim2023}. Using price histories, company profiles and news data, NASDAQ-100 returns  were predicted using LLMs in \cite{Yu2023} at weekly and monthly future points. Language models also generated explanatory narratives describing the ``chain of thought" reasoning behind the forecasts. Binary precision was reported as 69\% for one-month ahead predictions.

In \cite{Pasch2022}, an LLM was fine-tuned on news headlines, blogs and 10-K report data to predict one-year stock price direction at a 12 month investment horizon. Stocks were partitioned into three groups, based on price performance relative to the average price change of the entire market. Prediction of the group (\textit{good}, \textit{average} or \textit{bad}) was the target value in this scheme. Their findings reported an F1 score of 0.43 using news input data alone. In assessing the relative predictive value of different text sources as inputs to the model. Interestingly, they found that the 10-K reports were less valuable predictors than either news or blog articles \cite{Pasch2022}.

A final relevant work appears in \cite{Gupta2023}, where text data within annual reports from S\&P 500, 400 and 100 companies were encoded by an LLM. The model was prompted using natural language to answer queries regarding the forward-looking prospects of the company, based on the input text. LLM outputs in response were used as features to train a machine learning model (linear regression). The prediction variable was the percentage return of each stock between successive annual reports (12 months). In simulation, the highest top-\textit{k} ranked stocks were bought and returns estimated relative to the benchmark S\&P 500 index after one year. Cumulative returns from LLM picks outperformed the benchmark index over a time frame spanning years 2018 through 2023. No statistics of accuracy or F1 were shown in the results.

A general overview of LLMs can be found in \cite{Minaee2024}, and financial domain specific applications are reviewed in \cite{Nie2024}.

\section{Methods}

\paragraph{Problem definition.}
The specific problem addressed in this study is expressed as follows:
For a given company having adjusted closing stock price $p_{t}$
filing a 10-K report on current date $t$, predict the directional movement 
of the stock $D_{T}$
price expected at future date $T$:

\begin{equation}
	D_{T} =
	\begin{cases} 
		0, & \text{if $p_{t} \ge p_{t+T}$} \\
		1, & \text{if $p_{t} < p_{t+T}$} \\  
	\end{cases} \label{eqn-direction}
\end{equation}

where $T \in \{3, 6, 9, 12 \}$ months post-filing date, and $D_{T}=0$ corresponds to a \textit{sell} (or do not \textit{buy}) decision. The ability to make precise 
movement estimates of $D_{T}$ has obvious value for informing investment decisions
upon publication of a firm's 10-K report. This is a multiple-class learning problem, as both \textit{buy} and \textit{sell} are equally informative. 

\subsection{Data preparation}\label{sec:data-preparation}
An experimental data sample was constructed by downloading public SEC Form 10-K reports 
for years 2015-2024, for companies comprising the S\&P 500 index as of April, 2024. The companies
represented in this index cover around 80\% of available U.S. market capitalization; the index
itself provides a proxy for the overall stock market as it includes firms from all major
sectors of the U.S. economy.

Text was extracted from the narrative information in four major sections in each firm's structured report for the time period under study. These included: Item 1A: \textit{Risk Factors}; Item 3: \textit{Legal Proceedings}; Item 7: \textit{Management's Discussion and Analysis of Financial Condition and Results of Operations} (MD\&A); and Item 7a: \textit{Quantitative and Qualitative Disclosures about Market Risk}. Note that numerical data from Item 8: \textit{Financial Statements and Supplementary Data} were \textbf{not} used in these experiments. 

The raw data in these reports is voluminous and heterogeneous (text, image links, tables) embedded within XML. Various parsing difficulties during extraction were encountered for some reports. The final sample of companies for the current study numbered 477, of the nominal $\sim$ \!503 in the market index.

\paragraph{Summarization of 10-K text.}
Text from the four extracted items of interest were summarized in order to distill the essential concepts input to the stock movement forecast model. A processing pipeline to carry out this semantic refinement used an instruction-driven large language model (Mistral-7B \cite{Jiang2023}) and chatbot (ChatOllama) combined within the LangChain framework\footnote{\url{https://python.langchain.com/v0.1/docs/integrations/chat/ollama/}}.

\paragraph{Example construction.}
Examples for classification experiments were assembled by labeling each individual report with a list of future stock price movements post filing date for each prediction horizon, converted to a binary value as indicated in equation \ref{eqn-direction}. Individual data records  contained each company's unique SEC identification number, stock symbol, economic sector, report date, 10-K text and the target labels. This process created a dataset with multiple records for each company, one record per year in the experimental time frame.

The data were partitioned by ID number such that companies were not simultaneously represented in both training and out-of-sample test sets during the experiments. This was done to preclude possible information leakage between train and test data, as many company reports were anecdotally observed to contain highly similar sections of text across contiguous reporting years.

Train, validation and test data were grouped in approximate percentage ratios of 80:10:10, respectively.
Training examples were balanced by over-sampling the minority target class (\textit{sell}) \cite{Batista2004}.

\subsection{Experiments}
A small pre-trained language model\footnote{60M parameters, derived from Mistral\cite{Jiang2023};  \url{https://huggingface.co/typeof/mistral-60m}} was fine-tuned using the labeled 10-K example data, and trained to forecast directional stock price movements expected at discrete future points in time. 

Ten fine-tuning experimental trials were carried out for each  of 10 out-of-sample ``folds", and at each time horizon considered. Model weights were re-initialized to the pre-trained foundational model before each trial. A handful of passes through the training data were found sufficient to fine-tune the models at each data fold. Conventional machine learning statistics (F1-macro, precision, recall) \cite{Muslu2015}, \cite{Mitchell1997} were used to evaluate the out-of-sample performance of the forecasts, and aggregated statistics were compiled by averaging over the trials for each horizon.
In addition, performance by GICS economic sector was analyzed to identify variations in industry-specific predictability.

The smaller model was chosen for this application because its classification performance was found comparable to that of much larger models (e.g. Mistral-7B \cite{Jiang2023}), albeit with a much faster training time.

All experiments were carried out on a desktop gaming machine\footnote{Intel Core i7-12700K processor;  NVIDIA Tesla M40 and NVIDIA Titan XP GPUs; 32 GB DDR5-6000 RAM on-board}.

\section{Results and discussion}

Forecasting results for the out-of-sample test data appear in Tables \ref{tab:main-results} and \ref{tab:sectors-all-horizons}.

\paragraph{Aggregate performance.}
Overall classification performance at various prediction horizons is summarized in Table \ref{tab:main-results}. The statistics cover all economic sectors in the experimental sample. At each horizon, metrics for each  potential investment action are shown. F1-macro, Precision  and Recall values have been averaged over the 10 disjoint test data folds. The F1 score is the primary measure used to assess forecasting ability in this study, and is recommended for highly imbalanced data as indicated by  data shown in the Support column \cite{Narasimhan2016} of Table \ref{tab:main-results}. 

Two interesting observations can be made from the statistical results detailed in Table \ref{tab:main-results}. 

\textbf{First, the best predictive performance (as indicated by F1) is found at 6 and 9 months} (F1=0.62) after the 10-K report is published. Even at 12 months, F1 remains at a value of 0.59. This compares well to values reported elsewhere \cite{Kim2023} (F1=0.58 at 12 mo.), however those researchers utilized accounting data in addition to annual report text to generate price forecasts. In contrast, the current study is based entirely on  raw text input to the  language model.

\textbf{Second, the \textit{buy} signal is significantly more precise (compared to \textit{sell}) at all horizons} and can be acted upon with a degree of confidence for buy-and-hold investment. The best precision is seen at 6 months.  The model performs poorly when deciding that a stock price is expected to drop at the various time frames. Sensitivity of the \textit{buy} signal peaks at 9 months (recall=0.68), by a considerable amount relative to other prediction horizons.

It is somewhat surprising result that \textit{sell} signals were found imprecise and insensitive. Although the training data were balanced by artificially over-sampling the minority class (\textit{sell}), the test data samples were highly imbalanced. This may in part explain the objectively bad observed \textit{sell} signals in all cases considered. Another possibility is that there may be a propensity for managers to (consciously or otherwise) write the narrative sections of the 10-K report with a bias towards positive language, as suggested previously in \cite{Azimi2021}. Such practice would produce a sort of ``cognitive dissonance" and confuse the language model when being trained to predict a negative price movement based on largely positive forward-looking sentiments expressed in the report. 

\begin{table}[h]
	\centering
	\begin{tabular}{@{}llccccccc@{}}
		&                      & F1     & Precision 	& Recall 		& Support \\
		\toprule 
		3 mo.  & \textit{Sell} & 0.425  & 0.430     & 0.421 	& 1940 \\
		& \textit{Buy}  & 0.583  & 0.579     & 0.588  	& 2627    \\
		\hline 
		\vspace{-0.25cm}
		&     					&       	&      	&           &  \\
		6 mo.  & \textit{Sell} 	& 0.393  & 0.371     & 0.418  & 1649  \\
		& \textit{Buy}  	& \textbf{0.621}  	& \textbf{0.645}     & 0.599      & 2918 \\
		\hline
		\vspace{-0.25cm}
		&     					&       	&      	&           &  \\
		9 mo.  & \textit{Sell} & 0.406  & 0.467     & 0.360  & 2009      \\
		& \textit{Buy}  & \textbf{0.621}  & 0.574     & \textbf{0.677}  & 2558    \\
		\hline
		\vspace{-0.25cm}
		&     					&       	&      	&           &  \\
		12 mo. & \textit{Sell} & 0.462  & 0.453     & 0.471  & 1930    \\
		& \textit{Buy}  & 0.592   & 0.601     & 0.583  & 2637    \\
		\bottomrule
	\end{tabular}
	\caption{Aggregate performance results for out-of-sample test data. Statistics 
		compiled over all sectors and horizons. F1, precision and recall are calculated using ``macro" averaging. The best statistics are highlighted in bold.
	}
	\label{tab:main-results}
\end{table}

\paragraph{Comparison with random selection.}
The overall results of Table \ref{tab:main-results} are compared with a random decision to \textit{buy} or \textit{sell} stocks in Table \ref{tab:main-results-vs-rand}. The null hypothesis of ``no predictive value" in the 10-K text was tested by running 2500 trials using a pseudo-random number as a decision function. The F1 statistic is shown versus its randomized counterpart $\mathrm{F1_{rand}}$ for each horizon in the table. Signal above noise is given by the difference $\Delta = \mathrm{F1}-\mathrm{F1_{rand}}$.

The \textit{buy} signal of the language model provides between 4.8 and 9 percent improvement against a random selection, in an aggregate sense. The greatest differential  is seen at the 9 month horizon.

Contrarily, the \textit{sell} signal is worse than  na\"ive random choice at most  prediction  horizons. This result further corroborates the weakness of \textit{sell} signals generated by the models, as noted in the discussion  surrounding Table \ref{tab:main-results}.

\begin{table}[h]
	\centering
	\begin{tabular}{@{}llccc@{}}
		       &               & F1     & $\mathrm{F1_{rand}}$ & $\Delta$	\\
		\toprule
		3 mo.  & \textit{Sell} & 0.425  & 0.459   & -0.034\\ 
		       & \textit{Buy}  & 0.583  & 0.535   & 0.048\\ 
		\hline 
		\vspace{-0.25cm}
		       &     			&       &         & \\ 
		6 mo.  & \textit{Sell} 	& 0.393 & 0.419   & -0.026\\ 
		       & \textit{Buy}   & 0.621 & 0.561   & 0.060\\ 
		\hline
		\vspace{-0.25cm}
		       &     			&        &        & \\ 
		9 mo.  & \textit{Sell}  & 0.406  & 0.468  & -0.062\\ 
		       & \textit{Buy}   & 0.621  & 0.528  & \textbf{0.093}\\ 
		\hline
		\vspace{-0.25cm}
		       &     			&        &        & \\ 
		12 mo. & \textit{Sell}  & 0.462  & 0.458  & 0.004\\ 
		       & \textit{Buy}   & 0.592  & 0.536  & 0.056\\ 
		\bottomrule
	\end{tabular}
	\caption{Aggregate performance results versus random choice for out-of-sample test data, for 
		all sectors and horizons. F1 and $\mathrm{F1_{rand}}$  are calculated using ``macro" averaging.
	}
	\label{tab:main-results-vs-rand}
\end{table}

\paragraph{Performance by sector.}
Cross-sectional forecast performance results by economic sector are presented in Table \ref{tab:sectors-all-horizons}. The F1 columns represent scores obtained at different prediction horizons, in months after publication of the 10-K report. These data are general macro-averages and are not broken down by decision class.

The highest score overall is seen in Communication Services, observed at 12 months. The Materials sector at 9 month horizon shows the second best score. In the Energy group, the predictive value of annual report text is consistently good relative to other industries when viewed across the different horizons.
As seen by the horizon-averages (last row in the table), the 12 month horizon is associated with the best prediction results. This is in contrast with the aggregate F1 scores shown in Table \ref{tab:main-results}, where the 6 and 9 month time frames exhibit the best performance.

These results suggest that there are differences in narrative reporting styles across industries that are correlated with varying degrees and time scales of price movement predictability. 

\begin{table}[h!]
	\centering
	\begin{tabular}{rcccc}
		Sector                 & F1(3) & F1(6) & F1(9) & F1(12) \\
		\toprule
		Communication Services & 0.519 & 0.511 & 0.523 & \textbf{0.571}  \\
		Consumer Discretionary & 0.480 & 0.489 & 0.505 & \textbf{0.551} \\
		Energy                 & \textbf{0.521} & \textbf{0.552} & 0.511 & \textbf{0.550} \\
		Information Technology & 0.509 & 0.498 & \textbf{0.526} & 0.533  \\
		Health Care            & 0.473 & 0.460 & 0.477 & 0.524  \\
		Financials             & 0.504 & 0.509 & 0.518 & 0.516  \\
		Utilities              & 0.467 & \textbf{0.532} & \textbf{0.531} & 0.515  \\
		Consumer Staples       & \textbf{0.521} & 0.515 & 0.440 & 0.512  \\
		Industrials            & 0.513 & \textbf{0.516} & 0.518 & 0.509  \\
		Real Estate            & \textbf{0.541} & 0.514 & 0.516 & 0.508  \\
		Materials              & \textbf{0.539} & 0.493 & \textbf{0.565} & 0.495  \\
		\hline
		AVG.                   & 0.508       & 0.508    &  0.512         & 0.526  \\
		\bottomrule
	\end{tabular}
	\caption{F1 performance results by sector and prediction horizon (months, in parentheses). F1 is calculated using ``macro" averaging. The top three scores at each  horizon are highlighted in bold.}
	\label{tab:sectors-all-horizons}
\end{table}

\paragraph{General discussion.}

This study has demonstrated that the narrative text in a company's 10-K report has predictive utility for long-horizon stock price prediction. This finding is in contrast to results of previous work \cite{Pasch2022}, where annual reports were found to be less informative than news articles or even blogs at prediction of annual returns. Those authors \cite{Pasch2022} hypothesized that 10-K reports may be lacking in requisite information density to make accurate predictions at long horizons. The results of the current study suggest this is not the case.

The best predictions on price movements were seen at 6 and 9 months after publication of the 10-K.
A simple explanation for this result is that information contained in the report takes time to disseminate and be reflected in prices. This effect has been previously noted in \cite{Boguth2016}, where it was asserted that some types of information ``diffuse slowly into prices, often at different speeds for different securities". The findings reported here are consistent with this observation.

At the 12 month horizon, the F1-macro score obtained here is comparable to values reported elsewhere (current: F1=0.59; \cite{Kim2023}: F1=0.58). The present study did not include financial accounting data to augment the input space as in \cite{Kim2023}. It remains to be studied in future work whether or not such additional information  would improve the current statistical results in a material way.
	
As a final note, the experiments of this study were conducted using a relatively small language model, for reasons of compute-time efficiency and low cost. It is reasonable to speculate that forecasting results might improve with a larger model, following scaling laws studied in \cite{Kaplan2020}. Improvement is not guaranteed, at least for the current classification task. Smaller language models are subject to ongoing research, and be sufficient to obtain useful results in other applications \cite{Hillier2024}.

\section{Conclusion}

This paper documents a study in which small language models are fine-tuned to predict future stock price direction for S\&P 500 companies. The models are trained using data from several text-only sections of each firm's 10-K filings with the SEC. 

Experimental results demonstrate that precise ``buy" signals are generated on out-of-sample data for longer horizons, at 3, 6 and 12-month future points after publication of the report. Statistical predictive performance measures showed this method to be competitive with findings in previous comparative reports, without the presumed benefit of additional accounting data as input to a language model (e.g, \cite{Kim2023}).

The findings presented clearly show that much greater precision is possible for ``buy" versus ``sell" forecasts made by the language models. Sensitivity of the \textit{buy} signal peaks at 9 months versus the other prediction horizons.

As measured by the F1 score, the \textit{buy} signal provides between 4.8 and 9 percent improvement against a random stock selection model. 

In contrast, \textit{sell} signals generated by the models do not perform well. This may be attributed to the highly imbalanced out-of-sample data used in the experiments, or perhaps due to management drafting annual reports with a bias toward positive language as noted by previous researchers \cite{Azimi2021}.

An analysis of the relative performance by companies representing different economic sectors shows that the connection between 10-K text and forecastability of subsequent price movements varies by industry.

\bibliographystyle{abbrv}
\bibliography{./10K-LLM.bib}

\end{document}